\documentclass[twocolumn,showpacs,preprintnumbers,amsmath,amssymb]{revtex4}
\usepackage[english]{babel}
\usepackage{indentfirst}
\usepackage{graphicx}
\usepackage{amsmath}
\usepackage{amssymb}
\usepackage{amsthm}
\usepackage{mathrsfs}
\usepackage{lscape}
\usepackage{bm}
\usepackage{float}
\usepackage{longtable}
\usepackage{graphics}
\usepackage{color}

\begin{document}
\title{Confluence of resonant laser excitation and bi-directional quantum dot nuclear spin polarization}
\author{C. Latta$^1$, A. H\"ogele$^1$, Y. Zhao$^2$, A. N. Vamivakas$^2$, P. Maletinsky$^1$, M. Kroner$^1$, J. Dreiser$^1$, I. Carusotto$^{1,3}$, A. Badolato$^1$, D. Schuh$^1$,
W. Wegscheider$^1$, M. Atature$^2$, and A. Imamoglu$^1$}

\affiliation{$^1$ Institute of
  Quantum Electronics, ETH-Z\"urich, CH-8093 Z\"urich, Switzerland}
\affiliation{$^2$ Cavendish Laboratory, University of Cambridge, Cambridge CB3 0HE, UK} \affiliation{$^3$ BEC-CNR-INFM
and Dipartimento di Fisica, Universit\`a di Trento, I-38050 Povo, Italy}

\vspace{-3.5cm}

\date{\today}
\begin{abstract}
Resonant laser scattering along with photon correlation
measurements have established the atom-like character of quantum
dots. Here, we present measurements which challenge this
identification for a wide range of experimental parameters: the
absorption lineshapes that we measure at magnetic fields exceeding
1~Tesla indicate that the nuclear spins polarize by an amount that
ensures locking of the quantum dot resonances to the incident
laser frequency. In contrast to earlier experiments, this nuclear
spin polarization is bi-directional, allowing the electron+nuclear
spin system to track the changes in laser frequency dynamically on
both sides of the quantum dot resonance. Our measurements reveal
that the confluence of the laser excitation and nuclear spin
polarization suppresses the fluctuations in the resonant
absorption signal. A master equation analysis shows narrowing of
the nuclear Overhauser field variance, pointing to potential
applications in quantum information processing.

\end{abstract}
\pacs{} \maketitle

A number of ground-breaking experiments have demonstrated fundamental atom-like properties of quantum dots (QD), such
as photon antibunching \cite{Michler00} and radiative lifetime limited Lorentzian absorption lineshape
\cite{Hoegele04} of optical transitions. Successive experiments using transport \cite{Petta05} as well as optical
spectroscopy \cite{Marie05} however, revealed that the nature of hyperfine interactions in QDs is qualitatively
different than that of atoms: coupling of a single electron spin to the mesoscopic ensemble of $\sim 10^5$ QD nuclear
spins results in non-Markovian electron spin decoherence \cite{Loss02} and presents a major drawback for applications
in quantum information science. Nevertheless, it is still customary to refer to QDs as {\sl artificial atoms}; i.e.
two level emitters with an unconventional dephasing mechanism. Here, we present resonant absorption experiments
demonstrating that for a wide range of system parameters, such as the gate voltage, the length of the tunnel barrier
that separates the QDs from the back contact and the external magnetic field, it is impossible to isolate the optical
excitations of QD electronic states from a strong influence of nuclear spin physics. We determine that the striking
locking effect of any QD transition to an incident near-resonant laser, which we refer to as dragging, is associated
with dynamic nuclear spin polarization (DNSP); in stark contrast to previous experiments
\cite{Gammon01,Krebs06,Lai06,Koppens06,MaletinskyPRB07,Tartakovski07,ReillyScience08} the relevant nuclear spin
polarization is bi-directional and its orientation is determined simply by the sign of the excitation laser detuning.
We find that fluctuations in the QD transition energy, either naturally occurring \cite{Hoegele04} or introduced by
externally modulating the Stark field, are suppressed when the laser and the QD resonances are locked. We also find
that when the exchange interaction between the confined QD electron and the nearby electron Fermi-sea that leads to
spin-flip co-tunneling \cite{Smith} is sufficiently strong, it can suppress the confluence of laser and QD transition
energies by inducing fast nuclear spin depolarization \cite{MaletinskyPRL07}.

%%%%%%%%%%%%%%%%%%%%%%%%%%%%%%%%%%
\vspace{0.5 cm}

{\bf Locking of quantum dot resonances to an incident laser}

For a single-electron charged QD, the elementary optical excitations lead to the formation of trion states
($\mathrm{X}^{-}$) which are tagged by the angular momentum projection (pseudo-spin) of the optically generated
heavy-hole (Fig.~1A). In the absence of an external magnetic field ($B_{\rm{ext}} = 0$~T), the absorption lineshape
associated with these optical excitations are Lorentzian (Fig.~1B), with a linewidth $\Delta \nu \sim 2.0 \, \Gamma$;
here, $\Gamma \sim 1~\mu$eV is the spontaneous emission rate of the trion state. For $B_{\rm{ext}} > 100$~mT, the
coherent coupling between the (ground) electronic spin states induced by the transverse component of the
(quasi-static) nuclear Overhauser field is suppressed \cite{Atature06}: in this limit, the optical excitations of the
QD can be considered as forming two weakly coupled two-level-systems (Fig.~1A), where the blue (red) trion transition
is associated with a QD electron initially in state $|\uparrow_z \rangle$ ($|\downarrow_z \rangle$). Figure~1C shows
absorption measurements carried out at $B_{\rm{ext}} = 4.5$~T when the laser field with frequency $\omega_{\rm{L}}$ is
tuned across the blue-trion resonance (transition energy $\hbar \omega_{\rm{X}}$). For a laser that is tuned from an
initial blue-detuning to a final red-detuning with respect to the QD resonance (red line in Fig.~1C), we find that the
absorption is abruptly turned on when we reach $\omega_{\rm{L}} \sim \omega_{\rm{X}} + 2 \Delta \nu$; the absorption
strength then remains essentially fixed at its maximal value until we reach $\omega_{\rm{L}} \sim \omega_{\rm{X}} - 7
\Delta \nu$, where it abruptly goes to zero. A laser scan in the opposite direction (blue line) shows a complementary
picture with the trion absorption strength remaining close to its peak value for laser frequencies that are blue
detuned from the bare trion resonance by as much as $7 \Delta \nu$. The absorption scans of Fig.~1C show that the
trion resonance locks on to the laser frequency and can be dragged to either higher or lower energies by tuning the
laser frequency, provided that the scan frequency-step size is small. We observe this dragging effect for a wide range
of laser Rabi frequencies $\Omega_{\rm{L}}$ ranging from $\sim 0.3\, \Gamma$ to $3\, \Gamma$. We also emphasize that
dragging is not a simple line-broadening effect: the area of the absorption curve is an order of magnitude larger than
its $B_{\rm{ext}}=0$~T counterpart.

\begin{figure}[t]
\includegraphics[scale=0.7]{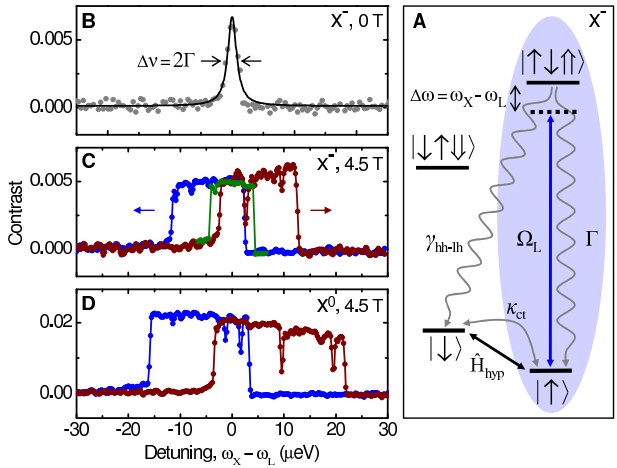}
\caption{\textbf{Figure 1(A) Dragging of quantum dot resonances.}Energy-level diagram of a single-electron charged
quantum dot: external magnetic field enables spin-selective excitation of Zeeman split trion (X$^{-}$) states.
Right-hand circularly polarized laser with Rabi frequency $\Omega_{\rm{L}}$ (blue arrow) and detuning $\Delta
\omega=\omega_{\mathrm{X}}-\omega_{\mathrm{L}}$ couples the higher energy Zeeman branch between the spin-up electron
ground state $|\uparrow\rangle$ and the trion state $|\uparrow \downarrow \Uparrow\rangle$. The relevant decoherence
processes (radiative decay $\Gamma$, co-tunneling $\kappa_{\rm{ct}}$ and heavy-hole/ligh-hole mixing
$\gamma_{\rm{hh-lh}}$) are indicated by grey arrows. (\textbf{B}), (\textbf{C}) Trion absorption spectra at zero
magnetic field (Lorentzian fit with a linewidth of $2~\mu$eV) and at $4.5$~T, respectively. For magnetic fields
exceeding $1$~T, the on-resonance scattering is maintained over many natural linewidths to both sides of the bare
resonance depending. Red (blue) spectra show data obtained by tuning the laser from an initial blue (red) to a final
red (blue) detuning from the quantum dot resonance. The green trace shows the energy range over which resonant
absorption is recovered in fixed laser detuning experiments (see Figure~3C). \textbf{(D)} The dragging effect is even
more prominent in the absorption spectra of the blue Zeeman branch of the neutral exciton X$^0$. In all experiments,
the temperature was 4.2~K.}
\end{figure}

It is well known that the optical response of a neutral QD for $B_{\rm{ext}} < 1$~T is qualitatively different, owing
to the role played by electron-hole exchange interaction \cite{Bayer99}. To assess the generality of the dragging
phenomenon, we investigated the response of a neutral QD for $B_{\rm{ext}}  \ge 2$~T: despite an energy level diagram
that is substantially different than that of a single-electron charged QD, the bright exciton transitions of a neutral
QD exhibit absorption lineshapes (Fig.~1D) that are qualitatively similar to that of a trion. In fact, we observe that
typical dragging widths for neutral QD exciton transitions are significantly larger than that of trion transitions
\footnote{Bi-directional DNSP is also observed when the red-trion or the red-Zeeman bright exciton transitions are
driven by a resonant laser field; in contrast to the blue-trion (Fig.~1C) and high-energy bright exciton (Fig.~1D)
transitions, the forward and backward scans in this case are highly asymmetric.}.

% Discussion around new Figure 2 (old Figure 3)

Further insight into the locking phenomenon depicted in Figs.~1C-D can be gained by studying its dependence on basic
system parameters. Figure~2 shows the two-dimensional (2D) map of resonant absorption as a function of laser frequency
and gate voltage $V_{\rm{g}}$ for two different sample structures exhibiting radically different ranges of the
co-tunneling rate. Fig.~2A shows the 2D absorption map of a QD that is separated from the Fermi sea by a 25~nm GaAs
barrier (sample~A): each horizontal cut is obtained by scanning the gate voltage for a fixed laser frequency. Red
(blue) bars show data obtained by scanning the gate voltage such that the detuning $\Delta
\omega=\omega_{\mathrm{X}}-\omega_{\mathrm{L}}$ decreases (increases). We estimate the co-tunneling rate
$\kappa_{\rm{ct}}$ for this sample at the center of the absorption plateau ($V_{\rm{g}} = 200$~mV) to be $1 \times
10^6 ~$s$^{-1}$ from electron spin pumping experiments carried out at $B_{\rm{ext}} = 0.3$~T \cite{Dreiser}. We
observe that the bi-directional dragging effect is strongest in the plateau center and is completely suppressed at the
edges ($V_{\rm{g}} \sim 160$~mV and $V_{\rm{g}} \sim 260$~mV).

\begin{figure}[t]
\includegraphics[scale=0.75]{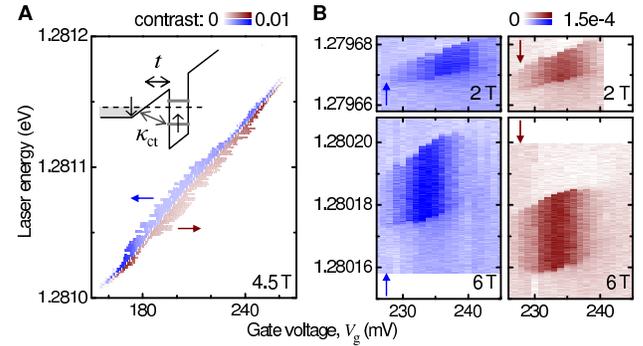}
\caption{\textbf{Figure 2 Dependence of dragging on system parameters.} Two dimensional absorption maps of the blue
trion Zeeman branch as a function of gate voltage and laser energy. Red (blue) arrows indicate the scan direction with
decreasing (increasing) laser detuning $\Delta \omega=\omega_{\mathrm{X}}-\omega_{\mathrm{L}}$. \textbf{(A)}
Absorption map recorded at $4.5$~T for a quantum dot in sample~A with a $25$~nm tunneling barrier: the data was
obtained by keeping the laser energy fixed and scanning the gate voltage. The inset illustrates the exchange coupling
to the Fermi reservoir that flips the quantum dot electron spin via co-tunneling events. The co-tunneling rate is
minimum in the center and maximum at the edges of the stability plateau; the dragging width scales inversely with the
co-tunneling rate $\kappa_{\rm{ct}}$. \textbf{(B)} Absorption maps of a quantum dot in sample B with a $43$~nm
tunneling barrier that were obtained by scanning the laser at a fixed gate voltage in external magnetic fields of
$2$~T (upper panel) and $6$~T (lower panel). Finite absorption contrast is restricted to the edge of the
single-electron charging plateau due to electron spin pumping. The set of data is complementary to \textbf{(A)} and
shows that dragging is independent of whether the laser or the gate voltage is scanned. The dragging width increases
sub-linearly with increasing strength of the external magnetic field.}
\end{figure}

Figure~2B shows absorption maps for a QD that is separated from the Fermi sea by a 43~nm tunnel barrier (sample~B) at
two different values of $B_{\rm{ext}}$ (2~T and 6~T): each vertical cut is obtained by scanning the laser energy for a
fixed gate voltage. The data presented in Fig.~2B is obtained for a narrow range of $V_{\rm{g}}$ near the edge of the
charging plateau: the large tunnel barrier drastically suppresses the tunnel coupling, such that the highest
co-tunneling rate (obtained at the plateau edges) coincides with the lowest rate obtained for sample~A \footnote{The
plateau middle sees a further $~6$ orders of magnitude reduction in the co-tunneling rate.}. Consequently,
bi-directional dragging in sample~B extends all the way out to the edge of the charging plateau, while in the plateau
center absorption disappears completely due to electron spin pumping into the $|\downarrow_z \rangle$ state
\cite{Atature06}. The overall range for dragging is $\sim 40~\mu$eV ($\sim 20~\mu$eV) for $B_{\rm{ext}} = 6$~T
($B_{\rm{ext}} = 2$~T) and is symmetrically centered at the bare X$^{-}$ transition energy. Further experiments
carried out on sample~C (not shown) with QDs separated by a 15~nm tunnel barrier from an electron reservoir and
exhibiting $\kappa_{\rm{ct}} > \Gamma$ throughout the plateau, did not show dragging effects. These results show that
locking of QD resonance to the incident laser energy is possible provided that the co-tunneling rate of the QD
electron satisfies $\kappa_{\rm{ct}} < 10^{8}~$s$^{-1}$ \footnote{We observed that the neutral exciton transition
(that is not limited by electron spin pumping) of sample~B QD exhibits dragging even at the center of the plateau
where we expect $\kappa_{\rm{ct}} \leq 1~ $s$^{-1}$. This observation suggests that dragging of the QD resonances
takes place even for QDs that are completely isolated from a back contact.}. Experiments carried out on QDs in all
samples showed that the dragging width increases sub-linearly with $B_{\rm{ext}}$ beginning at $\sim 1$~T.

% New discussion around new Figure 3

\begin{figure}[t]
\includegraphics[scale=0.75]{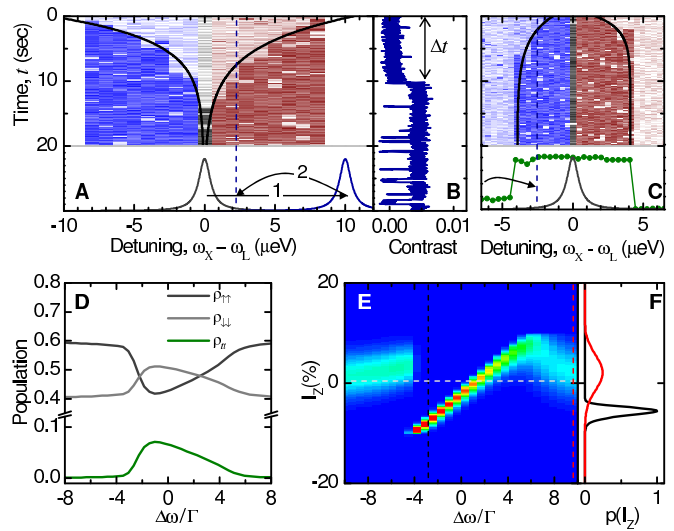}
\caption{\textbf{Figure 3(A) Bi-directional nuclear spin polarization.} Decay of nuclear spin polarization determined
by measuring the recovery of resonant absorption. The nuclei are dynamically polarized by voltage-controlled dragging
up to $10 ~\mu$eV, such that the resulting Overhauser field red (blue) shifts the trion transition, as indicated by
arrow 1 in the inset. A millisecond voltage ramp (arrow 2) takes the trion transition into a new detuning condition
$\mid \omega_{\rm{X}}-\omega_{\rm{L}} \mid\leq 8~\mu$eV and is accompanied by an instantaneous loss of the resonant
absorption contrast. The recovery of the full contrast after a time $\Delta t$ indicates that the nuclear spin
polarization has decayed to reestablish the resonant scattering locking condition. The decay is exponential (solid
curves in \textbf{A}) with decay constants $\tau_{\rm{d}} =3.7 \pm 0.7$~sec for blue and $4.9\pm0.9$~sec for red laser
detunings, respectively. \textbf{(B)} Time-trace of the on-resonance signal along the dashed line in \textbf{A},
demonstrating the bi-stable behavior of the absorption strength. \textbf{(C)} Buildup of nuclear spin polarization:
the nuclear spins are first depolarized by keeping the gate voltage in a region of fast co-tunneling. After
depolarization, a finite detuning of the laser from the trion transition is realized within milliseconds by abruptly
changing the gate voltage. The buildup time leading to maximum absorption contrast, associated with the build-up of
the Overhauser field, is exponential (solid lines with $\tau_{\rm{b}} =3.6\pm0.8$~sec for blue, and $2.1\pm0.7$~sec
for red laser detunings). \textbf{(D)} Simulation of the quantum dot level populations in steady-state for the
experiment described in \textbf{C}. \textbf{(E)} Probability distribution $p(I_{\rm{z}})$ for obtaining a value
$I_{\rm{z}}$ of the Overhauser field for a detuning range $(-10 \Delta \nu, \Delta \nu)$. The simulation shows
bidirectional nuclear spin polarization as well as a reduction in the Overhauser field variance when the system is
locked on to resonance. \textbf{(F)} A vertical line cut from \textbf{E} at zero (large) laser detuning plotted using
red (black) dashed line.}
\end{figure}

{\bf Bi-directional nuclear spin polarization}

The disappearance of dragging with increasing co-tunneling rate or vanishing external magnetic field strongly suggest
that locking of the QD optical transition to the laser frequency is associated with DNSP. Recent studies carried out
with laser fields resonant with the excited state transitions of QDs have shown that the nuclear spin depolarization
rate has a strong dependence on the electron spin co-tunneling rate \cite{MaletinskyPRL07}; these experiments also
demonstrated that when the QD is neutral, the nuclear spins do not depolarize even on time-scales exceeding an hour
\cite{Maletinsky09}. To confirm that DNSP indeed plays a key role in our experimental findings, we have carried out
experiments to determine the relevant time scales for the buildup and decay of the locking phenomenon in a regime of
minimum exchange coupling to the reservoir \footnote{A direct proof of DNSP using depolarization by resonant radio
frequency magnetic fields does not appear to be possible in self-assembled QDs  due to large inhomogeneous quadrupolar
fields \cite{Maletinsky09}.}.

Figure~3A presents a set of experiments that reveal the timescales associated with the decay of the DNSP generated
during dragging, for the trion transition in sample~A at $B_{\rm{ext}}=4.5$~T. The procedure used in these experiments
is to first drag the QD trion transition by about $5 \Delta \nu$ to the red side of the bare resonance
$\omega_{\rm{X}}$, and then to abruptly change the detuning between the trion and the laser field by a millisecond
voltage ramp. The effect of this ramp is to set a new detuning condition $\omega_{\rm{X}} - \omega_{\rm{L}} \leq 4
\Delta \nu$, which in turn results in an instantaneous loss of the absorption contrast. We observed that after a
waiting time on the order of seconds, the initially vanishing absorption strength recovered its maximum value
(Fig.~3A, red-coded data). By repeating this experiment for a set of final detunings ranging from $4 \Delta \nu$ to
$0$, we determined the characteristic exponential timescale for contrast recovery to be 5~sec. When we repeated the
same experiment by first dragging the trion resonance to the blue side of the bare resonance and monitoring absorption
for a final set of detunings satisfying $\omega_{\rm{X}} - \omega_{\rm{L}} \geq - 4 \Delta \nu$, we determined a
contrast recovery time of $\tau_{\rm{d}}=4~$sec (Fig.~3A, blue-coded data). These results provide information about
the decay time of the DNSP that is built up during the dragging process; as DNSP decays in the absence of a resonant
laser, the Zeeman shifted trion resonance frequency changes until it once again reaches resonance condition with the
laser field \footnote{Since the presence of a laser field that is detuned by less than $2\Delta \nu$ would speed up
the recovery of resonant absorption (see Fig.~3C), the timescales we obtain in this experiment could be regarded as an
upper bound on the nuclear spin depolarization rate.}. The DNSP decay times that we determine are in agreement with
the values one would extrapolate from earlier experiments where $\kappa_{\rm{ct}} \sim 10^8~$s$^{-1}$ resulted in DNSP
decay times on the order of few milliseconds at $B_{\rm{ext}} = 0.2$~T \cite{MaletinskyPRL07}. When we repeated the
experiment of Fig.~3 for the neutral QD exciton transition, we observed that the absorption contrast for
$|\omega_{\rm{X}} - \omega_{\rm{L}}| < 4 \Delta \nu$ always remained zero, indicating that the DNSP decay time was
much longer than our measurement time of $\sim 1$~hour; this observation is in perfect agreement with earlier
experiments \cite{Maletinsky09}. Finally, we remark that the bistable behavior of resonant absorption contrast that is
evident in vertical line cuts taken from the data in Fig.~3A (shown in Fig.~3B) is very characteristic of nonlinear
nuclear spin dynamics in QDs \cite{MaletinskyPRB07,Tartakovski07}.

Figures~1 and 2 demonstrate that the response of a QD to a given laser detuning strongly depends on how the system
reaches that particular detuning. To determine the QD optical response in the absence of such memory effects, we have
carried out another set of experiments, where we first set the laser frequency to a large negative detuning with
completely negligible excitation of the trion and kept the QD in a parameter regime with a strong co-tunneling rate
for several seconds. This procedure allows the QD nuclear spins to thermalize with the lattice, which in turn ensures
vanishing nuclear spin polarization. We then abruptly changed the voltage in milliseconds timescale, thereby
instantaneously establishing $|\omega_{\rm{X}} - \omega_{\rm{L}}| \le 7 \Delta \nu$. Subsequently, we observed the
time-dependence of the absorption signal at this fixed detuning. Figure~3C shows that for a detuning range of $ 3
\Delta \nu \ge |\omega_{\rm{X}}-\omega_{\rm{L}}| > \Delta \nu$, the absorption strength grows from zero to its maximum
value on a timescale of a few seconds while within $|\omega_{\rm{X}}-\omega_{\rm{L}}| \le  \Delta \nu$ the
on-resonance condition is established on a timescale below the temporal resolution limit of a few milliseconds in our
experiment. Even though the frequency range over which the QD is able to lock on to the laser field is identical for
red and blue detunings, the absorption recovery time is a factor of 2 slower for a laser that is tuned to the blue
side of the bare trion resonance. We also note that the frequency range over which full absorption is recovered in
these fixed laser frequency dragging experiments (green curve in Fig.~1C and Fig.~3C) is narrower than that of
dynamical dragging bandwidth obtained by tuning the laser across the resonance (red and blue curves in Fig.~1C).

%%%%%%%%%%%%%%%%%%%%%%%%%%%%%%%%%%%%%%%% Theory

Perhaps the most unexpected feature of our experiments is the remarkably symmetric dragging effect that  indicates
bi-directional DNSP; this observation is in stark contrast with recent experiments which demonstrated unidirectional
dragging of the electron (microwave) spin resonance and the bi-stability of the coupled QD electron-nuclei system
\cite{Vink09}. These results, obtained concurrently, could be understood as arising from a dominant nuclear spin
pumping process that is a nonlinear function of the degree of DNSP and competes with the nuclear spin depolarization
processes \cite{Rudner07,Nazarov08}.

We understand the bi-directionality of DNSP by considering that the optical excitation of the QD blue-trion transition
(Fig.~1A) induces two competing nuclear spin pumping processes that try to polarize QD nuclear spins in two opposite
directions and that have a different functional dependence on the laser detuning $\Delta \omega =
\tilde{\omega}_{\rm{X}} - \omega_{\rm{L}}$; here $\tilde{\omega}_{\rm{X}}$ is the QD blue trion transition energy
shifted by DNSP. As we argue below, the {\sl reverse-Overhauser} process \cite{Rudner07} that polarizes the nuclear
spins along $+\hat{z}$ direction depends linearly on the absorption rate $W_{\mathrm{abs}} = \omega_{\rm{L}}^2
\Gamma/(\Delta \omega^2 + \Gamma^2)$ from the $|\uparrow_{\mathrm{z}} \rangle$ state and dominates for large $|\Delta
\omega|$ \footnote{The reverse Overhauser process is associated with hyperfine assisted spin-flip Raman scattering and
is the resonant analog of the optical pumping effect that the early DNSP experiments \cite{Gammon01,Krebs06} were
based on.}. In contrast, the Overhauser process that polarizes the nuclear spins along $-\hat{z}$ scales as
$W_{\mathrm{abs}}^2$ and determines the direction of DNSP for small values of $|\Delta \omega|$ \footnote{The
Overhauser process is effected by optical pumping of the electron spin via heavy-light-hole mixing induced spontaneous
Raman scattering, together with a resonant-absorption induced spin dephasing mechanism that is proportional to
$W_{\mathrm{abs}}$. Since the excess spin population in state $|\downarrow_z \rangle$ is also proportional to
$W_{\mathrm{abs}}$, the overall spin-dephasing-induced Overhauser processes depend on $W_{\mathrm{abs}}^2$.}. The two
opposing processes balance each other out at a finite red or blue detuning \cite{Rudner07}. The red detuning
($\tilde{\omega}_{\rm{X}} > \omega_{\mathrm{L}}$) where the net polarization rate vanishes is a stable point of the
coupled system: tuning the laser closer to (away from) resonance would lead to an enhanced Overhauser
(reverse-Overhauser) process which would ensure that the trion resonance is blue (red) shifted until the stable point
is once again reached. In addition to the Overhauser and reverse-Overhauser channels, there are pure DNSP decay
processes that are induced either by exchange interactions with the Fermi sea or by electron-phonon interactions.

The interplay between the two polarization channels ensures that DNSP in either direction can be attained
\footnote{The unexpected dragging observed on the QD $X^0$ transition could also be understood as arising from a
similar competition between detuning dependent Overhauser and reverse-Overhauser processes. The asymmetry in
bi-directional dragging we observed in the red trion transition could be due to the fact that the (phonon mediated)
relaxation is reduced at low temperatures for the low energy trion state.}; the maximum DNSP that can be achieved is
in turn determined by the co-tunneling or phonon assisted DNSP decay processes. To gain further insight into this
interplay, we have solved the master equation describing the evolution of the coupled electron-nuclear spin system in
steady-state: to obtain a tractable master equation for $N \sim 1000$ spins, we have modeled the nuclear spin system
in the Dicke basis of collective spin states $|\bf{I},I_{\rm{z}}\rangle$ having a total angular momentum $\bf{I}$ ($0
\le I \le N/2$) and that couple to the QD electron via hyperfine interaction. States belonging to different $\bf{I}$
but having identical spin projection $I_{\rm{z}}$ along $B_{\mathrm{ext}}$ are assumed to be coupled weakly by
incoherent jump processes \footnote{This coupling would have been identically zero, if the hyperfine coupling to each
nuclei had been identical.}. The overall contribution of each $I$ to the coupled dynamics is weighted by the number
$D(I)$ of allowed permutation group quantum numbers associated with that total angular momentum \cite{Taylor03}. The
master equation is obtained by first applying a Schrieffer-Wolff transformation that eliminates the flip-flop terms of
the hyperfine interaction, and then tracing over the electron-spin and radiation field reservoirs in the Born-Markov
limit.

Figure~3D shows the steady state electron spin and trion populations evaluated for a range of fixed laser detunings
using this procedure. The trion population as a function of the laser detuning is a direct measure of the absorption
strength that we experimentally determined in Fig.~3C: we observe that our model reproduces the broadened absorption
profile but does not capture the sharp change in the absorption strength that we observe experimentally. This
discrepancy is most likely due to the fact that our calculations give the steady-state values which may be practically
impossible to observe experimentally. Fig.~3E shows the probability distribution $p(I_{\rm{z}}) = \sum_{\bf{I}}
\langle I_{\rm{z}} | \rho | I_{\rm{z}} \rangle$ for obtaining a specific value $I_{\rm{z}}$ of the Overhauser field,
evaluated for the same range of parameters used in Fig.~3C: these results demonstrate that the bi-directional dragging
indeed originates from a bi-directional DNSP. Perhaps more important is the fact that $p(I_{\rm{z}})$ reveals an
impressive reduction in the Overhauser field variance by more than a factor of 5, when dragging is present (Fig.~3F).

%%%%%%%%%%%%%%%%%%%%%%%%%%%%%%%%%%%%%%%%%%%%%%%%%%%%

% Discussion around Figure 4

\begin{figure}[t]
\includegraphics[scale=0.75]{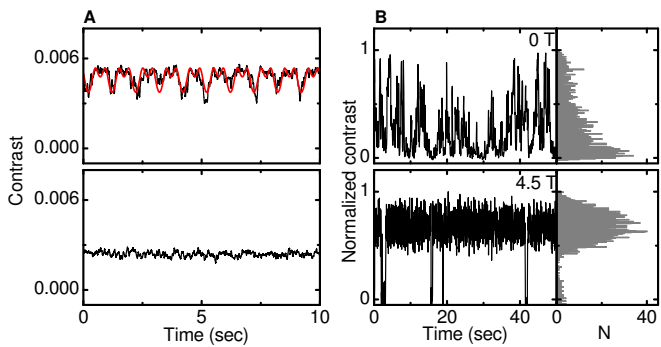}
\caption{\textbf{Figure 4(A) Suppression of fluctuations in quantum dot resonance frequency.} Suppression of the
fluctuations in the resonant absorption signal when the dragging condition is satisfied in sample B. Upper panel: the
measured absorption contrast as a function of time at a fixed laser at $B_{\rm{ext}}=0$~T in the presence of a 1 Hz
sinusoidal gate voltage modulation. The modulation amplitude corresponds to a shift of the trion transition by $\Delta
\nu/2$ starting from exact resonance condition. Lower panel: the corresponding time dependent fixed laser frequency
absorption contrast at $B_{\rm{ext}}=4$~T when dragging is present. The signal shows full compensation of the
externally induced detuning modulation. \textbf{(B)} The fluctuations in the resonant absorption signal observed for
the quantum dot in Sample A at $0$~T (upper panel) are suppressed in the presence of dragging at $4.5$~T (lower
panel); the data in both panels of \textbf{(B)} are normalized to peak absorption contrast obtained at the
corresponding magnetic field.}
\end{figure}

{\bf Suppression of fluctuations in quantum dot transition energy}

The experiments depicted in Figs.~1-3 as well as the numerical results shown in Fig.~3D demonstrate the existence of a
feedback mechanism that polarizes the necessary number of nuclear spins needed to shift the transition energy in a way
to maintain resonance with the excitation laser. Consequently, fluctuations in the QD transition energy that would
normally lead to a fluctuating absorption signal should be suppressed by such a compensation mechanism provided that
the fluctuations occur within the effective feedback bandwidth. Figure~4A shows the time-dependence of the absorption
signal for $B_{\rm{ext}}=0$~T and $B_{\rm{ext}}=4$~T for sample~B in the presence of external modulation of the trion
resonance energy. The upper panel shows a time record of the on-resonance signal for $B_{\rm{ext}}=0$~T  with
10~millisecond time resolution and in the presence of a controlled disturbance of the trion transition energy. This is
achieved by introducing a 1-Hz sine modulation on the gate voltage with a peak-to-peak voltage amplitude corresponding
to $ \Delta \nu/2$ via Stark shift. The $B_{\rm{ext}}=0$~T signal on resonance clearly reproduces this modulation as
the transition goes in and out of resonance with the laser consistent with the expected signal drop. However,
$B_{\rm{ext}}=4$~T on-resonance signal shows complete suppression of this external disturbance of the transition
energy (lower panel). The same experiment repeated for higher frequencies show that suppression works at least up to
10~Hz determining the relevant bandwidth for Overhauser field dynamics for this particular QD and gate voltage. When
the disturbance is in the form of a square wave modulation of peak-to-peak amplitude less than $\Delta \nu$,
suppression is still present, while beyond $\Delta \nu$ modulation, no suppression can be seen regardless of
modulation frequency. This is consistent with the slow timescale of DNSP buildup of sample B in comparison to the
abrupt jump of the transition under square wave modulation.

Even in the absence of an external perturbation, most QDs exhibit time-dependent fluctuations in $\omega_{\rm{X}}$
\cite{Hoegele04}: these fluctuations could arise either from the electromagnetic environment of the QD or fluctuating
nuclear Overhauser field. Figure~4B (upper panel) shows a typical time record of the resonant absorption signal of
sample A QD together with the corresponding distribution function for $\Omega_L \sim \Gamma$: in contrast to the
sample B QD studied in Fig.~4A, we observe up to $100 \%$ fluctuations in the resonant absorption signal, which is in
turn a factor of 3 larger than our noise floor. In contrast, for $B_{\rm{ext}} = 4.5$T (Fig.~4B lower panel), we find
that the fluctuations in the absorption signal are reduced to the noise-floor. The absorption signal occasionally
drops to zero, indicating bistability in the response of the coupled electron-nuclei system to the resonant laser
field. The frequency of jumps in absorption strength depends strongly on the system parameters; in particular ~sample
B QD practically never showed jumps during the observation period exceeding 10 sec.

%%%%%%%%%%%%%%%%%%%%%%%%%%%%%%%%%%%%
%Outlook

Having demonstrated that the locking of the QD resonance to the incident laser frequency via selective DNSP strongly
damps out the fluctuations in electronic transition energy, we address the possibility of suppressing the
time-dependent fluctuations in the nuclear Overhauser field \cite{ReillyScience08,Greilich06}. Given that the
effective Zeeman shift associated with the rms Overhauser field of the QD nuclei $B_{\rm{nuc}}$ is comparable to the
spontaneous emission rate, we would expect that the fluctuations in the Overhauser field would  lead to sizable
fluctuations of the resonant absorption signal on time-scales that are characteristic for the Overhauser field (few
seconds). Since we lack controlled experiments exclusively demonstrating the role of the Overhauser field fluctuations
on the absorption signal, we could only claim that the experiments depicted in Fig.~4 provide an indirect evidence for
a suppression of nuclear Overhauser field fluctuations that is predicted by our theoretical model (Fig.~3E and F).

We emphasize that a narrowing of the Overhauser field variance would have remarkable consequences for quantum
information processing based on spins. In particular, a major limitation for experiments detecting spin coherence in
QDs is the random, quasi-static Overhauser field that leads to an inhomogeneous broadening of the spin transition with
a short $T_2^*$ time \cite{Petta05}. Suppression of the long-timescale Overhauser field fluctuations by laser dragging
may be used to ensure that the time/ensemble averaged spin coherence measurements yield a dephasing time that is
limited only by the fundamental spin decoherence processes \cite{Coish04,Taylor07,Dassarma09}. While replacing the
more traditional spin-echo techniques with laser dragging would represent a practical advantage for optical
experiments, a more intriguing possibility would be the slowing down of nuclear spin dynamics by a combination of
large inhomogeneous quadrupolar shifts \cite{Maletinsky09} and dragging, which would in turn prolong the inherent
electron spin coherence time.

\begin{acknowledgments}

We thank Hakan T\"ureci, Jake Taylor, Geza Giedke, Mark Rudner and Lev Levitov  for discussions.

\end{acknowledgments}

\end{document}